\documentclass[prd, twocolumn, nofootinbib]{revtex4}

\usepackage{epsfig}

\newcommand{\beq}{\begin{equation}}
\newcommand{\eeq}{\end{equation}}
\newcommand{\beqa}{\begin{eqnarray}}
\newcommand{\eeqa}{\end{eqnarray}}

\def\pa{{\partial}}

\def\ha{\frac{1}{2}}

\begin{document} 

\title{Kinetic k-essence and Quintessence} 
\author{Roland de Putter \& Eric V.\ Linder} 
\affiliation{Berkeley Lab \& University of 
California, Berkeley, CA 94720} 

\date{\today} 

\begin{abstract} 
Dark energy models with non-canonical kinetic energy terms, k-essence, 
can have dynamical and sound speed properties distinct from canonical 
scalar fields, quintessence.  Concentrating on purely kinetic term 
Lagrangians, which can be technically natural, we investigate limits on 
the equation of state dynamics and sound speed behaviors and the extent 
to which these models can be separated from quintessence. 
\end{abstract} 

\maketitle

\section{Introduction} \label{sec:intro} 

In the quest for the physical origin of the cosmic acceleration, we have 
relatively little guidance from basic principles.  For dynamical scalar 
field models, e.g.\ quintessence, one must posit a potential, ideally 
possessing naturalness, 
without fine tuning.  However the potential receives quantum loop 
corrections from high energy physics, raising the problem of preservation 
of form, or technical naturalness.  An alternative approach is adopting 
non-canonical kinetic terms in the scalar field Lagrangian, leading to 
k-essence models.  

A subset of these -- kinetic k-essence -- offers the 
possibility of removing the potential altogether (or keeping it constant).  
With a shift symmetry in the field, $\phi(x)\to\phi(x)+\theta$, this 
provides technical naturalness, an important virtue for a physical theory, 
and so kinetic k-essence models are worth investigating.  Furthermore, 
these are in a sense as simple as quintessence in that they also involve a 
single function, here $\mathcal{L} = F(X)$ where $X$ is the kinetic energy. 

In this article we study the interplay between equations of state $w(a)$, 
defined in terms of an effective pressure to energy density ratio $w=p/\rho$ 
or equivalently cosmic expansion dynamics, and kinetic k-essence Lagrangians 
possessing certain stability properties. This will enable us to establish 
to what extent cosmological expansion or distance data could encounter 
degeneracies in the interpretation of the physical origin of the cosmic 
acceleration.  For example, what are the characteristics of a 
quintessence model appearing degenerate with a k-essence model, in the 
sense that they produce the same equation of state over some redshift 
range.  We derive limits on this degeneracy by showing which regions 
of the $w$-$w'$ phase space (where $w'=dw/d\ln a$, with $a$ the expansion 
scale factor) k-essence can lie 
in (see \cite{CaldLind05,Scherrer06,Linder07} for analyses of other DE 
models).  We will also consider the converse issue by going from some 
specific equations of state to the corresponding k-essence Lagrangians. 

In \S\ref{sec:motiv} we discuss some of the motivations for considering 
k-essence as a possible physical model for the origin of the cosmic 
acceleration.  \S\ref{sec:model} presents a pedagogic overview and 
explanation of some of the main properties of (purely kinetic) k-essence. 
In \S\ref{sec:stability}, we analyze the stability of solutions and from 
this derive the condition the DE equation of state must satisfy 
for it to be degenerate with a stable kinetic k-essence model. 
In \S\ref{sec:reconstruction}, we give the closed form solution for 
this degeneracy condition and present a simple prescription for 
predicting the behavior of any kinetic k-essence Lagrangian.  
In \S\ref{sec:examples} we exhibit the results for several illustrative 
examples. 

\section{Motivation} \label{sec:motiv}

About a decade ago, analysis of the luminosity-redshift relation of 
distant supernovae \cite{Perl99,Riess98} suggested that the 
expansion of the universe is accelerating. This has led people to 
postulate the existence of a mysterious new component to the universe 
called dark energy (DE), characterized by a negative equation of state 
$w = p/\rho$. Since then, many observations of different kinds, most 
notably of the cosmic microwave background \cite{Sperg06} and large 
scale structure \cite{Teg04}, have consolidated the picture of a 
spatially flat universe with contributions to the critical energy 
density of about 1/4 by matter (dark matter plus baryonic matter) and 
3/4 by dark energy. One of the biggest challenges in modern physics has 
become to explain the nature of this dark energy.

A standard explanation of DE is Einstein's cosmological constant $\Lambda$, 
a constant uniform background energy density corresponding to an equation 
of state $w = -1$. The physical origin of $\Lambda$ could perhaps be a 
vacuum energy from particle physics. A natural value for this vacuum energy density would be either of order $\rho_{\rm vac} \approx M_{\rm fund}^4$, where $M_{\rm fund}$ is some fundamental mass/energy scale in nature like the Planck scale $M_p$ or perhaps the electroweak scale $M_{EW}$, or exactly zero if some symmetry (like supersymmetry) enforces it. However, the dark energy density is about 
3/4 of the current critical energy density of the universe, giving $\rho_{\Lambda} \approx 10^{-122} M_p^4 \approx 10^{-54} M_{EW}^4$, which is outrageously small. This problem of the dark energy being so small (but not zero) compared to the natural physical scales is called the Cosmological Constant Problem.

Another fine-tuning problem related to the cosmological constant model is the fact that the dark energy and matter components are comparable in size today. If the dark energy truly is a cosmological constant, there is only a short period of time in the evolution of the universe when 
this is the case so why is it happening exactly now, while we are around to observe it? This issue is part of the so called Coincidence Problem.

Even though a cosmological constant is in good agreement with the current data, the problems above provide ample motivation to look for DE models beyond a cosmological constant. The most popular alternatives to $\Lambda$ are scalar field models, in particular quintessence (see \cite{CopSamTsu06,Linder07} 
and references therein), which describes a single scalar field $\phi$ with a standard Lagrangian density $\mathcal{L} = X - V(\phi)$, with $X = \ha \pa_{\mu} \phi \pa^{\mu} \phi$. Quintessence models can describe a range of equations of state necessary for an accelerated expansion, but like $\Lambda$ suffer from fine-tuning issues.

More recently, scalar field models with non-canonical kinetic energy have gained interest. These so called k-essence (the ``k'' standing for kinetic) models are described by Lagrangians of the general form \cite{Melchetal03} 
$\mathcal{L} = v(\phi) F(X) - 
V(\phi)$ (a canonical kinetic energy is given by $F(X)=X$ and $v(\phi) = 1$). 
K-essence was originally proposed as a model for inflation 
\cite{PicDamMukh99}, and then as a model for dark energy 
\cite{ArMukhStein00}, along with explorations of unifying dark energy and 
dark matter \cite{BilTupViol02,BenBerSen02}.  It now appears increasingly likely from 
both theoretical stability issues 
and observational constraints (e.g.\ \cite{Sandetal04,BeanDore03,AmWagFin05}) from 
matter clustering properties (dark matter is very clumpy while DE is quite 
smooth out to the Hubble scale) that dark matter and dark energy are not 
the same substance; we will treat k-essence purely as a dark energy 
candidate. One reason for the interest in k-essence is that it admits 
solutions that track the equation of state of the dominant type of matter 
(in the early universe this is radiation) until pressure-less matter 
becomes dominant, at which point the k-essence begins to evolve toward 
cosmological constant behavior 
\cite{ChiOkYa00,ArMukhStein00,ArMukhStein01}. Such behavior can to a 
certain degree solve the fine-tuning problems mentioned above. 

A good way to look at k-essence is as a generalization of canonical 
scalar field models (i.e.\ quintessence).  Let us consider the quintessence 
Lagrangian in a bit more detail. Where the fact that it must be a 
function only of $X$ and $\phi$ comes from Lorentz invariance, the 
reason for the kinetic energy term being equal to $X$ (i.e.\ canonical) 
is merely that we assume $X$ to be small compared to some energy scale 
and higher order terms to be irrelevant. Even though this is often a correct assumption because of the Hubble damping, there exist cases where it is not. In k-essence, we look at models where the higher order terms are not necessarily negligible. This can give rise to interesting new dynamics not possible in quintessence.

Another motivation for studying k-essence is the relation of scalar field theory to the quantum mechanics of a single particle (see also \cite{BagJassPad03}). Heuristically, field theory can be viewed as the continuum limit of a grid of particles, with the field at a certain point describing the excitation of the particle at that point. The canonical scalar field theory Lagrangian density can in that picture be seen as the generalization of the Lagrangian of a non-relativistic point particle $L = \ha m \dot{q}^2$ (where $q$ is the particle's position). The Lagrangian of a relativistic point particle $L = - m \sqrt{1 - \dot{q}^2}$ on the other hand leads to a non-canonical field theory Lagrangian density $\mathcal{L} = - \sqrt{1 - 2X}$, i.e.\ a k-essence Lagrangian.

A third motivation for the study of k-essence is that non-canonical Lagrangians appear naturally in string theory. In particular, the tachyon effective Lagrangian (\cite{Sen02_2}, also see \cite{Sen04}, in particular section 8, for a very readable review) has the Dirac-Born-Infeld-like form $\mathcal{L} = - V(\phi) \sqrt{1 - X}$. Here, the field $\phi$ represents the tachyon condensate describing the evaporation of a D-brane. The Nambu-Goto action for a $p$-brane embedded in a $p+2$ dimensional space-time can be written in the same form \cite{Goratal04}, except that $\phi$ has the role of the coordinate transverse to the brane in this scenario. We discuss this in a little more detail in the Appendix.

Finally, as mentioned in the Introduction, technical naturalness as from a 
shift symmetry gives an advantage for purely kinetic k-essence Lagrangians, 
and these involve only a single function, $\mathcal{L} = F(X)$, like 
quintessence.

\section{The model} \label{sec:model} 

We study k-essence, dark energy described by a single, real scalar 
field $\phi$, minimally coupled but with a non-canonical kinetic term.  
In general, the k-essence action is of the form 
\begin{equation}
\label{full action}
S = \int d^4x \sqrt{-g}\, F(\phi,X),
\end{equation}
where $X := \ha \pa_{\mu} \phi\, \pa^{\mu} \phi$.  We concentrate on 
the subclass of kinetic k-essence, with a $\phi$-independent action 
\begin{equation}
\label{action}
S = \int d^4x \sqrt{-g}\, F(X).
\end{equation}
We assume a Friedmann-Robertson-Walker metric $ds^2 = dt^2 - a^2(t)\, d\vec{x}^2$ (where $a(t)$ is the scale factor) and work in units $c = \hbar = 1$. Unless explicitly stated otherwise, we assume $\phi$ to be smooth on scales of 
interest so that $X = \ha \dot{\phi}^2$. Note that this implies $X \geq 0$.

Varying the action (\ref{action}) with respect to the metric gives the energy momentum tensor of the k-essence
\begin{equation}
\label{en_mom}
T^{\mu \nu} = F_X \pa^{\mu} \phi\, \pa^{\nu} \phi - g^{\mu \nu} F,
\end{equation}
where a subscripted $X$ denotes differentiation with respect to $X$. Using that for a comoving perfect fluid the energy momentum tensor is given by $T_{\mu \nu} = - p g_{\mu \nu} + (\rho + p) \delta_{\mu}^0 \delta_{\nu}^0$, the k-essence energy density $\rho$ and pressure $p$ are
\begin{equation}
\label{en}
\rho = 2 X F_X - F
\end{equation}
and
\begin{equation}
\label{mom}
p = F.
\end{equation}
Throughout this paper, we will assume that the energy density is positive so that $2 X F_X - F > 0$. The equation of state is
\begin{equation}
\label{eq state}
w = \frac{p}{\rho} = \frac{F}{2 X F_X - F}.
\end{equation}

The equation of motion for the field can be found either by applying the Euler-Lagrange equation for the field to the action (\ref{action}), or by plugging the energy density and pressure given above into the continuity equation for a perfect fluid. Either way, the result is
\begin{equation}
\label{EoM1}
F_X \ddot{\phi} + F_{XX} \dot{\phi}^2 \ddot{\phi} + 3 H F_X \dot{\phi} = 0,
\end{equation}
or equivalently, in terms of $X$,
\begin{equation}
\label{EoM2}
(F_X + 2 F_{XX} X) \dot{X} + 6 H F_X X = 0,
\end{equation}
where a dot denotes differentiation with respect to $t$ and $H = \dot{a}/a$ is the Hubble parameter.
This equation can be integrated to give
\begin{equation}
\label{X sol}
X F_X^2 = k a^{-6},
\end{equation}
with $k \geq 0$ a constant \cite{Scherrer04}. 

Note that equation (\ref{X sol}) tells us that the possible solutions $X(a)$, and therefore the behavior of all physical properties of the k-essence (like $\rho$, $p$ and $w$) {\it as a function of the scale factor}, are completely determined by the function $F(X)$ and do not depend on the evolution of the other types of 
energy density. The only dependence of the k-essence component on other 
components enters through $a(t)$.  One consequence of this is to preclude 
the possibility of tracking solutions \cite{ZlatWangStein99} that automatically 
follow the equation of state of the dominant form of matter in the universe.  
Tracking behavior {\it is} possible in general k-essence models that do 
have $\phi$-dependence in the action \cite{ChiOkYa00,ArMukhStein00,ArMukhStein01}. 

An interesting distinction when discussing dark energy models is between 
dark energies with $w > -1$ and those with $w < -1$. The latter are 
referred to as phantom dark energy \cite{Caldwell02} and can have 
rather exotic properties. For instance, their energy density is an 
increasing function of the scale factor, which can be seen from the 
Friedmann equation $d\ln \rho/d\ln a = - 3 (1 + w)$ (see \cite{AbrPin06} for a discussion of problems arising in phantom k-essence theories). The boundary 
between the phantom and non-phantom regime is $w=-1$, e.g.\ a time 
independent cosmological constant. If a DE evolves from one regime to 
another, this is called phantom crossing, but \cite{Sen06}, e.g., showed 
that this is impossible for a purely kinetic k-essence. We refer to 
\cite{Vik05,CaldDor05} for a discussion of phantom crossing in 
the context of other DE models.

For kinetic k-essence, one can use equation (\ref{eq state}) to express 
the condition $w > -1$ ($w < -1$) as a condition on the function $F(X)$. 
We need to consider the two possibilities $F > 0$ and $F < 0$ separately. 
In the first case, demanding the energy density be positive immediately 
implies $w > 0$.  For $F < 0$, a positive energy density means that 
$2 X F_X/F < 1$ so 
\beq
\label{}
w = \frac{-1}{1 - 2 X F_X/F} > -1
\eeq 
when $F_X > 0$.
All together, the conditions become (cf.\ \cite{AbrPin06})
\beqa 
\label{fconditions1}
F>0 &\Longrightarrow& w>0, \\ 
F<0\quad \& \quad F_X > 0 &\Longrightarrow& w>-1, \label{fconditions2}\\ 
F<0 \quad \&\quad F_X < 0 &\Longrightarrow& w<-1. \label{fconditions3}
\eeqa 
(Recall the condition $\rho>0$, or $F_X>F/(2X)$, is implicit). 

We conclude this section by pointing out a useful scaling 
property of the kinetic k-essence Lagrangian, namely $F(X)\to CF(BX)$ 
with $B>0$ and $C$ arbitrary constants, leaves the physical properties, 
such as equation of state $w(a)$, unchanged. 
In other words, once we have found an $F(X)$ that reproduces an equation 
of state of interest, we are free to rescale both $X$ and $F$ without 
affecting $w(a)$. The freedom to rescale $F$ by a factor $C$ follows 
from the fact that both $p$ and $\rho$ are proportional to $F$; $C$ could 
play the role of a constant potential (see \S\ref{subsec:4}).  The freedom to 
rescale $X$ comes from the fact that one can always redefine the field 
$\phi \to \phi/\sqrt{B}$ without changing the physics. The reason we 
mention this property is because in the following we will often use it 
to rescale $F(X)$ into a convenient form (for example with $X = 1$ and 
$F = - 1$ at redshift zero) or to leave out multiplicative constants in 
expressions for $X$ or $F$. 

\section{Restrictions on $w(a)$ from Stability}
\label{sec:stability}

In this section we restrict possible equations of state by demanding 
that the k-essence be stable against spatial perturbations. Since the 
k-essence action only depends on $X$ and not on $\phi$, the relevant quantity for determining whether or not a solution of the equation of motion (\ref{X sol}) is stable is the adiabatic sound speed squared
\beq
\label{sound X}
c_s^2 := \frac{p_X}{\rho_X} = \frac{F_X}{2XF_{XX} + F_X} = \frac{F_X^2}{(X F_X^2)_X}.
\eeq
Perturbations can become unstable if the sound speed is imaginary, $c_s^2<0$, 
so we insist on $c_s^2 > 0$, or equivalently \cite{AbrPin06} 
\beq
\label{}
(X F_X^2)_X > 0.
\eeq

\begin{figure}
  \begin{center}{
  \includegraphics*[width=8.5cm]{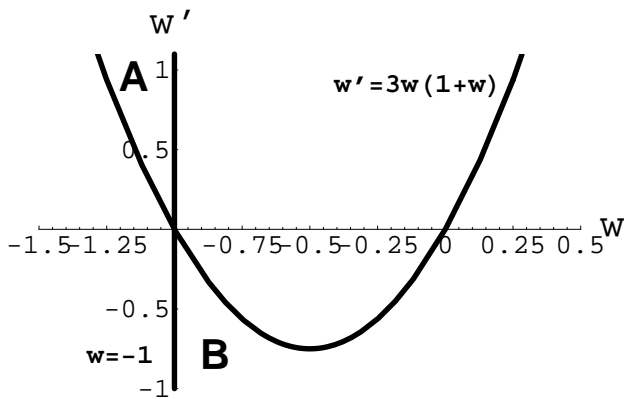}
  }
  \end{center}
  \caption{The requirement that $c_s^2 > 0$ restricts the equation of state to the two regions A and B, bounded by the lines $w=-1$ and $w'=3w(1+w)$. }
  \label{regions}
\end{figure}

Writing the condition $c_s^2 > 0$ in terms of $w(a)$, using $c_s^2 = (dp/da)/(d\rho/da)$ (which is valid as long as $dX/da \neq 0$), places a restriction on the equations of state $w(a)$ that can be described by stable k-essence solutions. Using
\beq
\label{drho da}
\frac{d\rho}{da} = - \frac{3 (1 + w)}{a} \rho
\eeq
and
\beq
\label{dp da}
\frac{dp}{da} = \frac{dw}{da} \rho + w \frac{d\rho}{da} = \frac{a (dw/da) - 3 w (1 + w)}{a} \rho,
\eeq
we get
\beq
\label{restriction}
c_s^2 = \frac{dp/da}{d\rho/da} = \frac{3 w (1 + w) - w'}{3 (1 + w)} > 0,
\eeq 
where $w' := dw/d\ln a$. This restricts the equation of state to lie within two regions of the $w - w'$ plane (see Figure~\ref{regions}), bounded by the lines $w = -1$ and $w' = 3 w (1 + w)$. The first line separates phantom k-essence ($w < -1$) from ordinary k-essence and the second is the constant pressure line, as can be seen from Eq. (\ref{dp da}).  As mentioned previously, it is not possible to cross 
between regions A and B \cite{Sen06}. As we will show later, the requirement that $w(a)$ lies in region A or B implies that a number of popular ansatzes for $w(a)$ cannot realistically describe (stable) k-essence.  

Note that a combination of stable kinetic k-essence models stays in 
the stable region \cite{Linder07} (and a combination of unstable models 
stays unstable).  However, $\Lambda$ plus a matter 
component (or more generally a $w\ge0$ component; also see \cite{HolNai04}) 
can look like stable k-essence. 
Another potentially interesting requirement to consider is $c_s^2 \leq 1$, which says that the sound speed should not exceed the speed of light, which suggests violation of causality. This condition would add an extra line to 
Figure~\ref{regions}, given by
\beq
\label{null}
w' = -3 (1 - w^2).
\eeq
However, we will not impose this condition because even though $c_s^2 > 1$ means signals can travel faster than light, this does not appear to lead to causal paradoxes (see for example \cite{Brun07}).

\section{Distinguishing kinetic k-essence from quintessence}
\label{sec:reconstruction}

If we know which equations of state can be reproduced by kinetic k-essence, we may be able to distinguish it observationally from other DE candidates. In the previous section, we derived a necessary condition on the DE equation of state for it to be described by a stable k-essence solution.  
In this section, we will answer a different question, namely what equations of state can be described by k-essence in the first place? We will find that condition (\ref{restriction}) from the previous section is in fact a sufficient condition. Any equation of state that stays within region A or B of the $w - w'$ plane can in principle be described by a k-essence Lagrangian density $F(X)$. Moreover, we will show how to construct this function. Our method consists of finding $F$ and $X$ as a function of the scale factor by expressing them in terms of physical quantities that can in principle be measured. These functions can then be used to construct $F(X)$. The first part is well defined for any dark energy, i.e.\ we can always formally construct $X(a)$ and $F(a)$ once we know, say, $w(a)$. However, if the slope of $X(a)$ changes sign, one ends up with a double valued function $F(X)$. In those cases, the dark energy cannot be described by a well defined k-essence model. We will see that double valuedness is produced exactly when crossing one of the lines $w = -1$ or $w' = 3 w (1 + w)$ in the $w-w'$ plane.

Now let us find $F(a)$ and $X(a)$. Equation (\ref{mom}) tells us that $F$ is simply equal to the pressure
 \beq
\label{F(a) 1}
F(a) = p(a).
\eeq
Inserting this into equation (\ref{eq state}) and solving for $X F_X$ gives
\beq
\label{}
X F_X = \frac{1 + w}{2 w} p,
\eeq
which squares to
\beq
\label{}
X (X F_X^2) = \left(\frac{1 + w}{2 w}\right)^2 p^2.
\eeq
Upon insertion of equation (\ref{X sol}), this leads to
\beq
\label{X(a) 1}
X(a) = C a^6 (\rho(a) + p(a))^2,
\eeq
where $C > 0$ is a constant we can freely choose.  
Equations~(\ref{F(a) 1}) and (\ref{X(a) 1}) define effective quantities 
for $F$ and $X$ even if there is no actual k-essence.  
If a well defined k-essence Lagrangian does exist, $F$ and $X$ have 
their standard meaning as the Lagrangian and as $\frac{1}{2} \pa_{\mu} 
\phi \pa^{\mu} \phi$ respectively. 

It is useful to find an expression for $X(a)$ in terms of the equation of state $w(a)$. Unfortunately, since $\rho$ and $p$ are in general integrals of $w(a)$, it is not possible to find a closed expression. However, it is possible to construct a differential equation. Differentiating 
equation~(\ref{X(a) 1}) with respect to $a$ and using 
equation~(\ref{drho da}) gives 
\beq
\label{diff X}
\frac{dX}{d\ln a} = - 6 \left(\frac{3 w (1 + w) - w'}{3 (1 + w)}\right) X = - 6 c_s^2 X.
\eeq
For $F(a)$, we have the equation
\beq
\label{diff F}
\frac{dF}{d\ln a} = - \frac{3 w (1 + w) - w'}{w} F.
\eeq
Given some evolution of the equation of state, we can derive $X(a)$ using 
Eq.~(\ref{diff X}) and then use Eq.~(\ref{diff F}) to find $F(X)$. 
Alternatively, inverting the $X(a)$ from Eq.~(\ref{diff X}), one can 
use Eqs.~(\ref{eq state}) and (\ref{X sol}) to get 
\beq
\label{F using X}
F(X) \propto \frac{w(a(X))}{a^3(X) (1 + w(a(X)))} \sqrt{X}.
\eeq

From the dynamics in the $w$-$w'$ plane, one can predict what sort of 
k-essence solution this corresponds to.  If the dynamics crosses a 
boundary defining the four regions in Fig.~\ref{regions}, then $c_s^2$ 
changes sign, indicating a pathology within the k-essence picture.  
Note from Eq.~(\ref{diff X}) that $c_s^2$ changing sign 
corresponds precisely to the slope of $X(a)$ changing sign, and at the 
same time $F(X)$ becomes double valued.  However, 
any equation of state that stays within one of the four regions can in 
principle be obtained from a k-essence Lagrangian. The stability argument 
from Section \ref{sec:stability} selects regions A and B from those four 
regions.   In conclusion, imposing stability, k-essence can generate 
precisely the equations of state that lie in regions A and B of the 
$w$-$w'$ plane.  

Conversely, one can look at a given $F(X)$ and determine whether it is 
a viable k-essence model and what a corresponding quintessence model 
would be like.  (Note that \cite{Maletal03} considered a similar question 
in terms of the effective quintessence potential.)  
First, from the slope of the function one can deduce by 
applying Eqs.~(\ref{fconditions1})-(\ref{fconditions3}) whether the model is 
phantom or not.  Since the k-essence field cannot cross $w=-1$ then the 
function $F(X)$ cannot change the sign of its slope, and it cannot be 
double valued (requiring the slope to go infinite).  From the curvature of 
the function (concave or convex), in combination with the slope, one can 
read off whether the adiabatic sound speed is real or imaginary, and hence 
look for stability: 
\beqa 
w>-1&:&\quad F_{XX}>-\frac{F_X}{2X}, \quad [{\rm suff.\ } F_{XX}\ge0]\, \label{fxxp}\\ 
w<-1&:&\quad F_{XX}<-\frac{F_X}{2X}, \quad [{\rm suff.\ } F_{XX}\le0]\,. \label{fxxm}
\eeqa 
Here suff.\ indicates a sufficient (but not necessary) condition for 
stability; this also corresponds to $0<c_s^2\le1$.  

Since by eye one can 
usually tell when the curvature is convex or concave, the sufficient 
condition can be a useful guide.  However, if $F_{XX}$ is near but on 
the wrong side of zero, then one must calculate the value to establish 
stability (although in any case the adiabatic sound speed would exceed 
the speed of light in these ambiguous cases).  One can apply 
Eqs.~(\ref{fxxp})-(\ref{fxxm}) to the $F(X)$ plots in the examples of 
the following section to verify their usefulness as a quick indicator 
of the stability of possible kinetic k-essence Lagrangians. 

To clarify further possible degeneracies between k-essence and quintessence 
physics, we examine in the next section specific examples that illustrate 
the points made above.

\section{Examples}
\label{sec:examples}

We here consider a few particular ansatzes for the DE equation of state 
$w(a)$ and analyze their degree of degeneracy with a (stable) k-essence 
solution.  To begin, we check whether the equation of state lies today ($z=0$) 
in one of the allowed regions given by Eq.~(\ref{restriction}). If so, a 
function $F(X)$ can be constructed to reproduce this equation of state in 
a certain redshift range around $z = 0$. If $w(a)$ stays in the allowed 
region for all $a$, then we can find an $F(X)$ degenerate with this 
dynamics for all redshifts.  (But it is important to note that it will 
not be equivalent to a quintessence field because the sound speed will not 
be unity; the two different physical models will be distinguishable to the 
extent that spatial inhomogeneities in the dark energy component are 
relevant.)  

\begin{figure}
  \begin{center}{
  \includegraphics*[width=8.5cm]{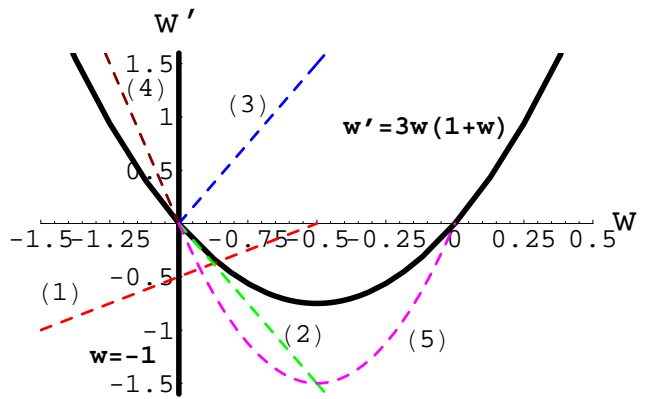}
  }
  \end{center}
  \caption{We investigate for a number of equations of state whether or not they can be described by a kinetic k-essence model. The plot shows the trajectories in the $w - w'$ plane for some examples. It also shows the boundaries between stable and unstable k-essence regions (see Fig.~\ref{regions}): the lines $w = -1$ and $w' = 3 w (1 + w)$. The equations of state are: (1) $w = -0.9 + 0.4 (1-a)$, see \S\ref{subsec:2}; (2) $1+w = (1 - 0.9) a^{-3}$, (3) $1+w = (1 - 0.9) a^{3}$, (4) $1+w = (1 - 1.1) a^{-6}$, all \S\ref{subsec:3}; (5) the Chaplygin gas, see \S\ref{subsec:4}. Note that $w = {\rm const}$ equations of state (see \S\ref{subsec:1}) correspond to points on the horizontal axis.} 
  \label{regions2}
\end{figure}

If $w(a)$ leaves the allowed regions after some time, then we can only 
match stable k-essence with quintessence over the redshift range where 
$w(a)$ does lie in one of the allowed regions. Since in practice we can 
only measure $w(a)$ in a limited redshift range anyway, the relevant 
question is whether we can reproduce a given equation of state in the 
redshift range constrained well by data, not necessarily for all 
redshifts.  For each sample equation of state considered below, we 
establish the acceptable redshift range and evaluate the equivalent 
Lagrangian function $F(X)$ and sound speed $c_s^2$. The examples are designed to
lie in different regions of the phase  
space with different stability properties, for illustration; see Fig.~ 
\ref{regions2}.

\subsection{Constant $w$}
\label{subsec:1}

A constant equation of state is simple and approximates some well 
known components.  For example, 
radiation has $w  = +1/3$, matter $w = 0$ and a cosmological constant 
$w = -1$.  Moreover, a (canonical) free field theory, with zero 
self-interaction potential $V=0$, where the Lagrangian is simply 
$F(\phi,X)=X$, has $w=+1$.  

It is useful to discuss the cases $w = -1$ and $w \neq -1$ separately.  
From Eq.~(\ref{eq state}), there are two ways to obtain $w = -1$. The 
first is by having $X =$ const. Since constant $X$ implies constant 
$\rho$, such solutions can {\it only} give $w = -1$. From 
Eq.~(\ref{X sol}), $X =$ const is a solution if $F_X = 0$. These 
solutions are special because since $dX/da = 0$ we cannot use 
Eq.~(\ref{restriction}) for the sound speed; instead we have to go back 
to Eq.~(\ref{sound X}). We see that (under the assumption that $F_{XX} 
\neq 0$, otherwise we have a canonical scalar field) this case has 
$c_s^2 = 0$ and is thus marginally stable. If 
$X$ is not constant, the only way to get $w \equiv -1$ is by having 
$F =$ const. This is just a cosmological constant and the adiabatic 
sound speed is not defined because there are no perturbations. 

\begin{figure}
  \begin{center}{
  \includegraphics*[width=8.5cm]{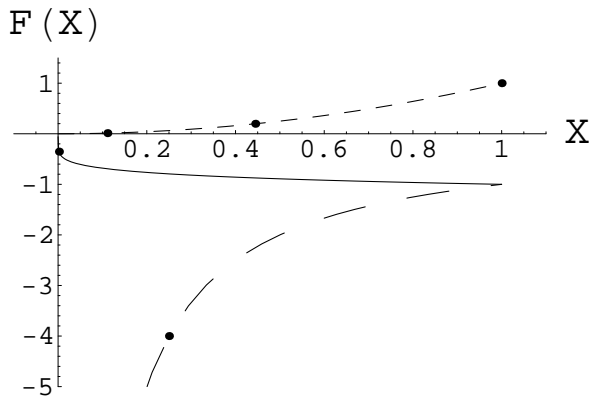}
  }
  \end{center}
  \caption{$F(X)$ for constant $w$ (ansatz A) with $w = - 1.5$ (solid), 
$w = -1/3$ (long dashed) and $w = +1/3$ (short dashed). For $w<0$, we 
choose $X(z=0) = 1$ to be at the end of the plotted domain; the field 
evolves with time from left to right and the marked points correspond 
to $z = 1$. For $w>0$, the field moves toward $X=0$ and the markers 
on the $w=+1/3$ curve indicate from left to right $z = 0, 1, 2$. The 
Lagrangians for negative constant $w$ ($\ne -1$) correspond to unstable 
solutions. An equation of state $w \equiv -1$ can be obtained from any $F(X)$ with an extremum by letting $X$ sit at that extremum (or from a cosmological constant $F =$ const).}
  \label{ansatz1}
\end{figure}

Next we look at $w \neq -1$. 
One can obtain constant $w \neq -1$ solutions if $X$ is not constant and the functional form of $F(X)$ has the right form. Since constant equations of state are points on the horizontal axis in figure \ref{regions}, we need $w \geq 0$ for the equation of state to be in the stable region.  More quantitatively, 
from Eq.~(\ref{restriction}), we have
\beq
\label{}
c_s^2 = w.
\eeq
In other words, any solution leading to a constant negative equation of state ($\neq -1$) is unstable viewed as k-essence. 

It is straightforward to explicitly construct k-essence Lagrangian 
densities corresponding to constant $w$ without applying the machinery 
developed in the previous section (see also \cite{ChimFein04}). Instead, we can just use 
Eq.~(\ref{eq state}) to get 
\beq
X F_X = \frac{1 + w}{2 w} F,
\eeq
which integrates to
\beq
\label{const w sol}
F(X) \propto X^{\frac{1 + w}{2 w}}.
\eeq
Since the energy density should be positive, $F$ must be positive for the $w > 0$ solutions and negative for the $w < 0$ ones.  As discussed in 
\S\ref{sec:model}, we are free to choose the magnitude of the proportionality factor.

Let us now look at some specific cases. First of all, it is useful to note that we cannot reproduce matter-like behavior, as for $w = 0$ the k-essence Lagrangian (\ref{const w sol}) is not well defined. This result makes sense: $w = 0$ corresponds to a pressure-less material, whereas in the case of k-essence, the pressure is the same as the Lagrangian density. In other words, $w = 0$ would mean that the Lagrangian density is zero everywhere, which of course just means there is no DE model at all.

Equation (\ref{const w sol}) also confirms that $w = 1$ corresponds to a canonical free field (no potential) Lagrangian $F(X) = X$ (which is simultaneously a k-essence and a quintessence model).  Note that skating models (see 
\cite{Linder07} and references therein), moving along a constant potential 
with $w'=-3(1-w^2)$, stretch between true free fields $V=0$, $w=1$ and 
pure constant potentials $V=V_0$, $w=-1$, following $X\sim a^{-6}$.  
This leads by Eq.~(\ref{X sol}) to $F_X={\rm const}$, or $F\sim X+ 
{\rm const}$.  
In this sense one can think of kinetic k-essence models as skaters that 
``push off'', altering their kinetic energy.  
Radiation-like behavior, $w=1/3$, is generated by a function of the 
form $F(X)=X^2$.  
(Note this should not be interpreted 
as a perturbation around a minimum of a potential since the field motion 
does not correspond to rolling in $F(X)$.)  
Figure~\ref{ansatz1} illustrates several examples.

\subsection{Time variation: $w(a) = w_0 + w_a (1 - a)$}
\label{subsec:2}

\begin{figure}
  \begin{center}{
  \includegraphics*[width=8.5cm]{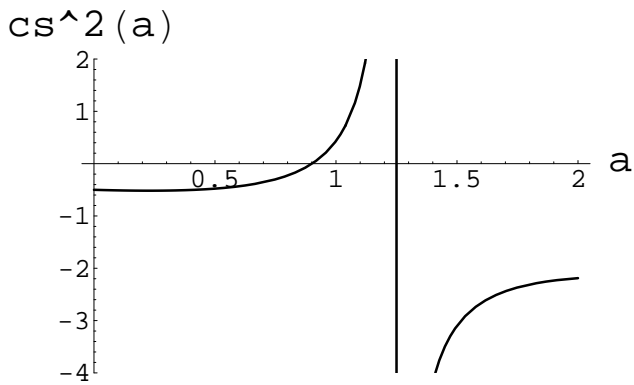}
  }
  \end{center}
  \caption{Ansatz B with $(w_0,w_a) = (-0.9, 0.4)$.  The adiabatic sound 
speed squared $c_s^2$ is plotted vs.\ the scale factor $a$, clearly 
showing that $c_s^2 > 0$ today but becomes 
negative both at $a \approx 0.9$ and $a \approx 1.25$. Hence, this 
equation of state can only be described by a stable kinetic k-essence 
solution in the (rather limited) range between those times (also see 
Fig.~\ref{ansatz2 2}).  }
  \label{ansatz2}
\end{figure}

A commonly used ansatz allowing for time variation is $w(a)=w_0+w_a(1-a)$, 
which provides a good fit to many canonical scalar field and other model 
behaviors \cite{Linder03}, at least over the past expansion history. 
Since $w' = - w_a a = w - (w_0 + w_a)$, in the $w$-$w'$ plane the 
equation of state starts in the past somewhere on the $w$-axis (at 
$w(a\ll1) = w_0 + w_a$).  Thus from the previous subsection we know it 
cannot be wholly degenerate with a stable k-essence model.  Its dynamics 
corresponds to a straight line with slope one that may cross through 
the stable region but again in the far future lie in an unstable region. 
Thus there is only a finite range of redshift when it may look like a 
stable kinetic k-essence model. 
The values of $w$ and $w'$ today are given by $w(a = 1) = w_0$ and 
$w'(a = 1) = - w_a$.  For a model within the stable region today we 
consider $(w_0,w_a) = (-0.9,0.4)$, in Fig.~\ref{ansatz2}.  This crosses 
the $w' = 3 w (1 + w)$ line at $a \approx 0.9$. This means that when 
calculating $F(X)$, starting at $a = 1$, it will become double valued 
at $a \approx 0.9$. In the future, the equation of state crosses the 
other boundary $w = - 1$ at $a \approx 1.25$. Hence, this equation of state 
is only degenerate with a stable kinetic k-essence model in the very 
limited range $a \approx 0.9 - 1.25$.

\begin{figure}
  \begin{center}{
  \includegraphics*[width=8cm]{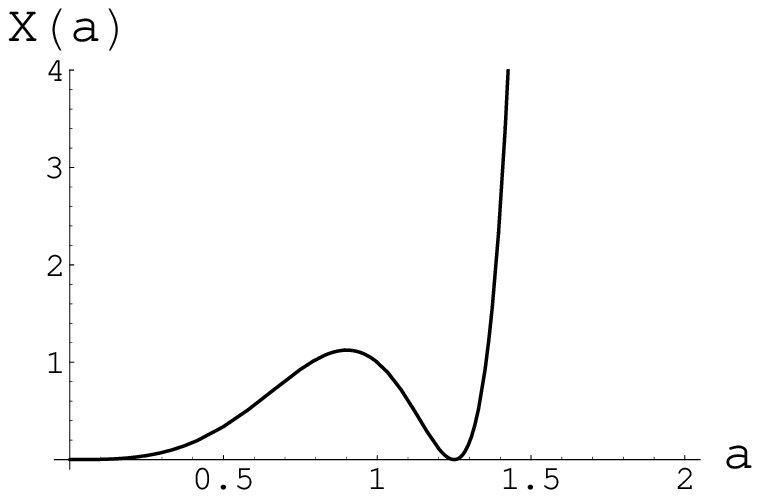}
  \includegraphics*[width=8cm]{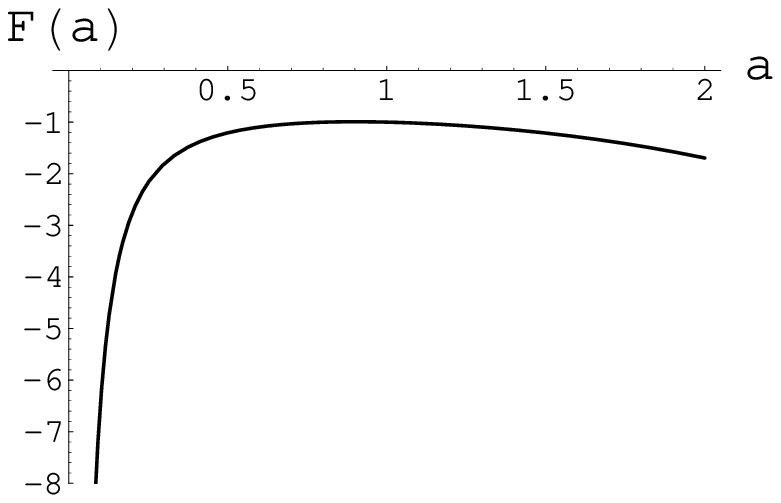}
  \includegraphics*[width=8cm]{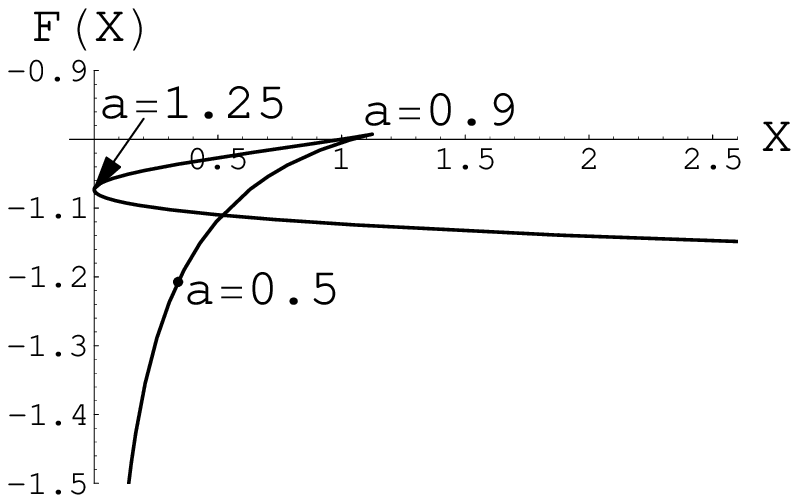}
  }
  \end{center}
  \caption{As Fig.~\ref{ansatz2}, but 
plotting $X(a)$ and $F(a)$ (see text and Fig.~\ref{ansatz2} for details), 
normalized such that today $X = 1$ and $F = - 1$. The third panel shows 
the function $F(X)$ obtained from $X(a)$ and $F(a)$ over the range 
$a = 0.1 - 1.5$. $F(X)$ turns around at the points where $dX/da = 0$ 
(i.e.\ where $c_s^2$ becomes negative, at $a \approx 0.9$ and $a \approx 
1.25$) and is therefore not single valued. Hence, this $(w_0,w_a) = 
(-0.9, 0.4)$ example model only corresponds to a well defined 
$F(X)$ in the region $a \approx 0.9- 1.25$. 
}
  \label{ansatz2 2}
\end{figure}

To calculate the k-essence Lagrangian explicitly, we need to solve 
Eq.~(\ref{diff X}). This can be done analytically in the case at hand, giving
\begin{equation}
\label{}
X(a) = C a^{- 6 (w_0 + w_a)} [1 + w_0 + w_a (1 - a)]^2 e^{6 w_a a}\,, 
\end{equation}
where $C$ is a positive constant. 
In the region between its extrema (the points where $w$ crosses one of 
the boundaries in Fig.~\ref{regions}), $X(a)$ can be uniquely inverted 
and $F(X)$ can be computed after using Eq.~(\ref{diff F}) to find $F(a)$.  
The solution is shown in Fig.~\ref{ansatz2 2} for $(w_0,w_a) = (-0.9,0.4)$.  
Note how $F(X)$ becomes double valued when going beyond the extrema.

\subsection{Thawing/Freezing Regions} 
\label{subsec:3}

The ansatz of $w'=x(1+w)$ leads to an equation of state
\beq
1+w(a)=(1+w_0)a^x.
\eeq
This describes a thawing model for $x>0$, where the equation of 
state starts frozen at high redshift so $w=-1$, and then begins rolling 
away from it.  This form describes well a number of renormalizable 
power law potentials and pseudo-Nambu Goldstone boson (PNGB) models. 
For $x<0$ it is simply a toy model of an equation of state approaching 
(freezing into) a cosmological constant state. See \cite{CaldLind05} for more
on thawing and freezing models. 

From Eq.~(\ref{restriction}), the adiabatic sound speed for these models 
is given by
\beq
\label{}
c_s^2(a) = w(a) - \frac{x}{3}.
\eeq
This shows that $c_s^2$ is positive today only if $x < 3 w_0$. It is positive at all times if $w_0 > -1$ and $x < -3$ (for $w_0 < -1$, it can only lie in the stable region for a limited redshift range regardless of the value of $x$). Still, we see that the criterion that $c_s^2$ be positive already rules out all the thawing equations of state and part of the freezing ones.

To construct $F(X)$, we again solve Eq.~(\ref{diff X}), giving 
\beq
\label{X sol ans3}
X(a) = C a^{2x + 6} e^{-\frac{6 (1+w_0)}{x} a^x}\,, 
\eeq 
where $C$ is a positive constant. 
Note that if $(x+3)/(1+w_0)>0$ this function has an extremum at 
$a = \{(x+3)/[3(1+w_0)]\}^{1/x}$. After finding $F(a)$ from 
Eq.~(\ref{diff F}), we can again calculate $F(X)$.  

\begin{figure}
  \begin{center}{
  \includegraphics*[width=8.5cm]{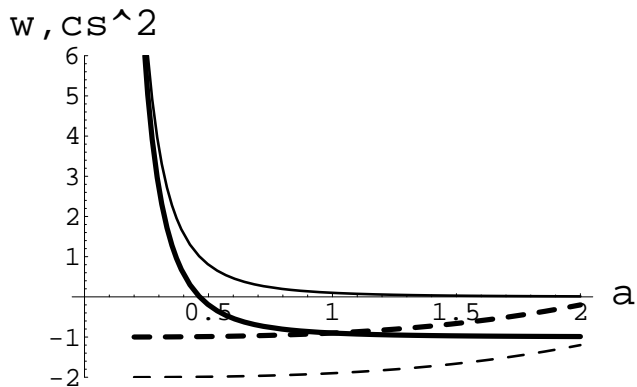}
  }
  \end{center}
  \caption{Ansatz C with $w_0=-0.9$ and $x = -3$ (solid) or $x = 3$ 
(dashed).  The equation of state (thick lines) and adiabatic 
sound speed squared (thin lines) are plotted vs.\ $a$. 
Note $x = 3$ lies in a forbidden region ($c_s^2 < 0$) for all $a$ and 
therefore cannot correspond to a stable kinetic k-essence solution. 
For $x = -3$, the corresponding k-essence Lagrangian is exhibited in 
Fig.~\ref{ansatz3 2}.} 
  \label{ansatz3}
\end{figure}

\begin{figure}
  \begin{center}{
  \includegraphics*[width=8.5cm]{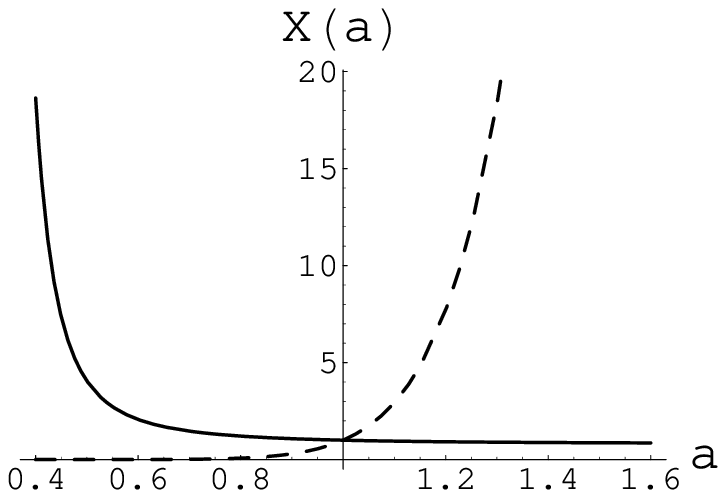}
  \includegraphics*[width=8.5cm]{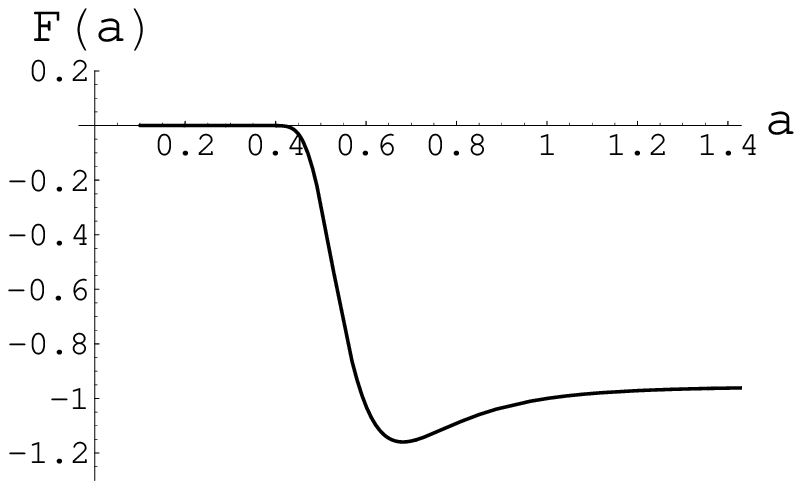}
  \includegraphics*[width=8.5cm]{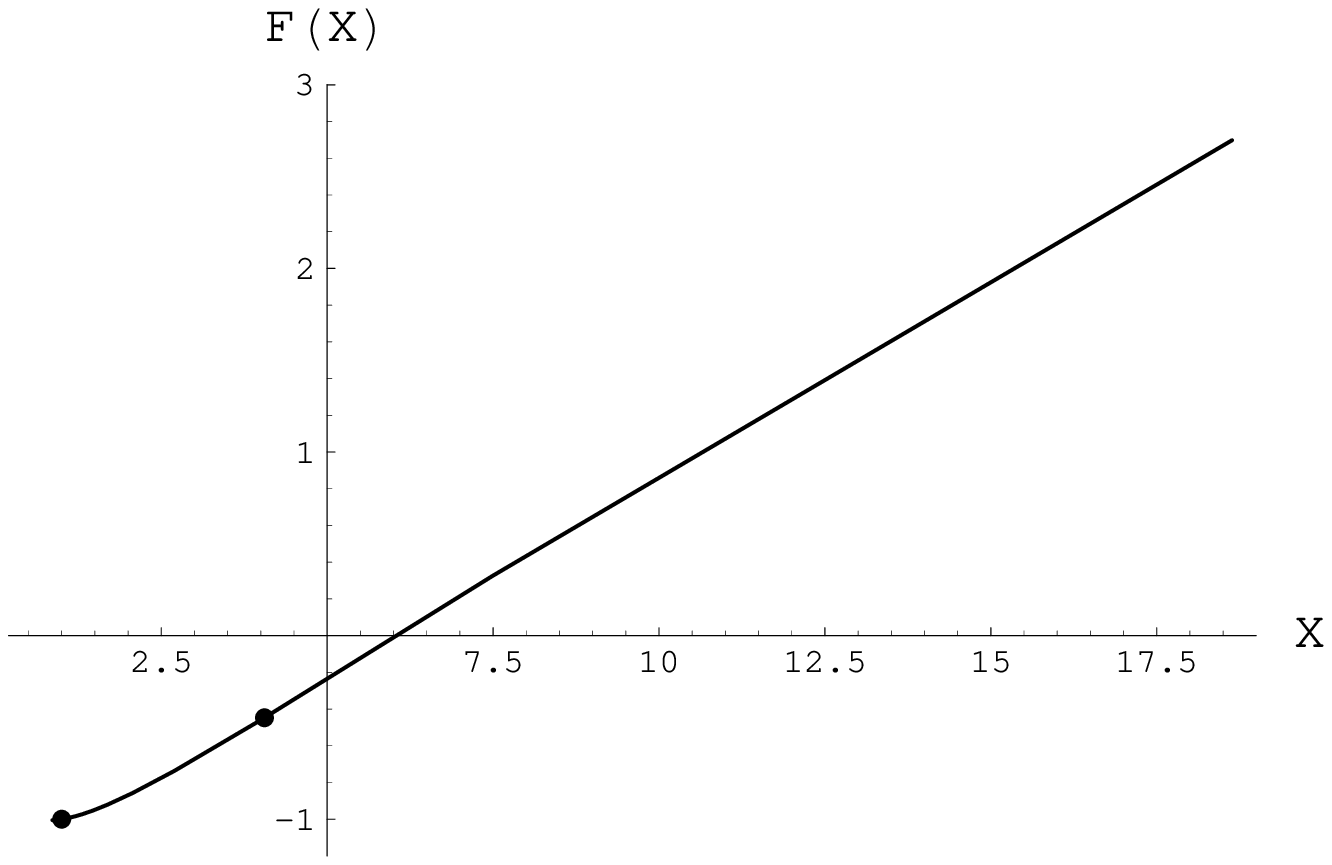}
  }
  \end{center}
  \caption{
As Fig.~\ref{ansatz3} but showing the k-essence functions $X(a)$, $F(a)$, 
and $F(X)$.  In the last panel $F(X)$ is plotted for the stable $x=-3$ case 
over $a = 0.4 - 1.6$ (from top right to bottom left), with markers at $z=1$ 
(right) and $z=0$ (left). } 
  \label{ansatz3 2}
\end{figure}

\begin{figure}
  \begin{center}{
  \includegraphics*[width=8.5cm]{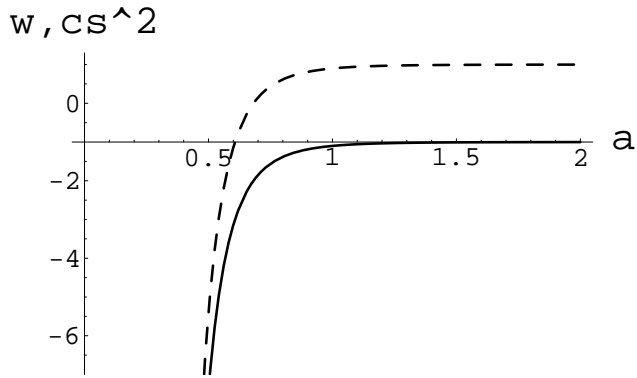}
  }
  \end{center} 
  \caption{As Fig.~\ref{ansatz3}, but for the phantom case $w_0 = -1.1$ 
and $x=-6$.  Here the solid curve shows $w(a)$ and the dashed curve $c_s^2$.}
  \label{ansatz3 phantom}
\end{figure}

\begin{figure}
  \begin{center}{
  \includegraphics*[width=8.5cm]{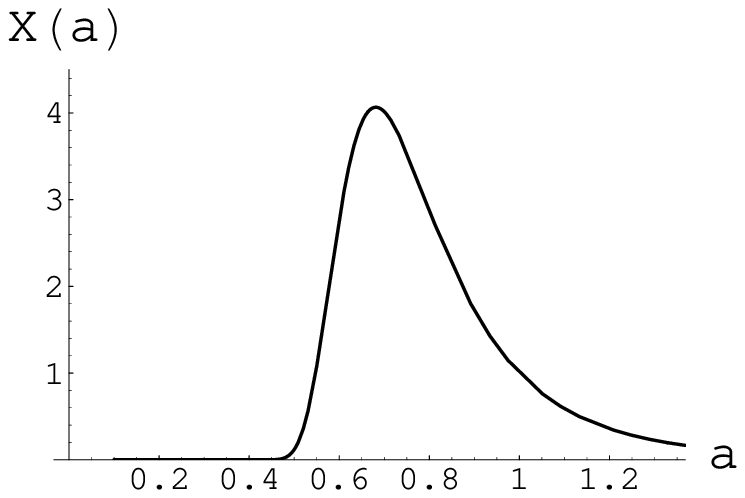}
  \includegraphics*[width=8.5cm]{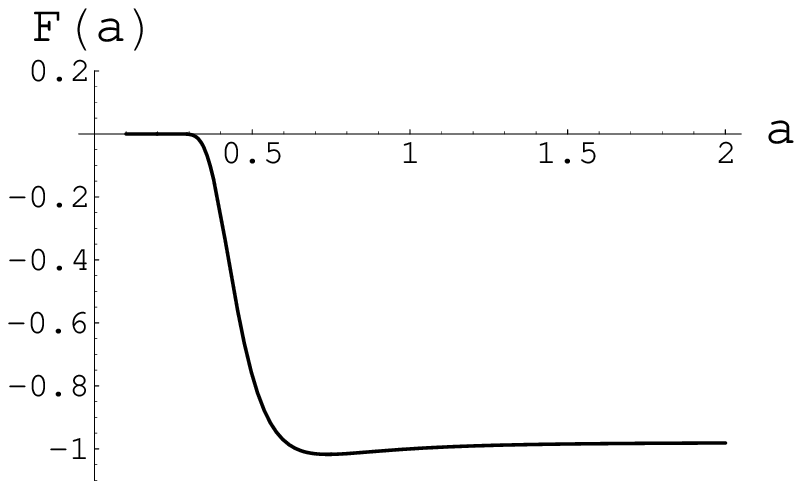}
  \includegraphics*[width=8.5cm]{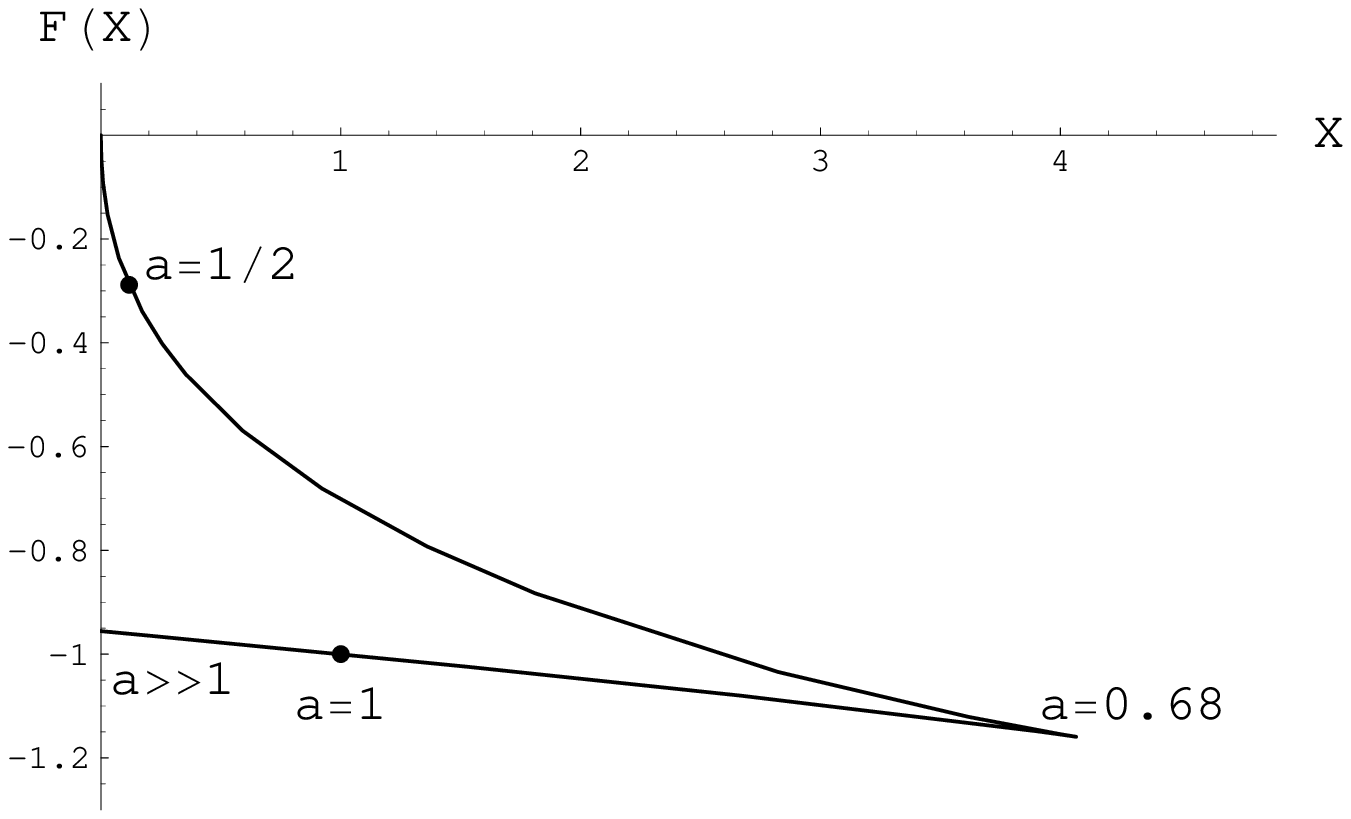}
  }
  \end{center}
  \caption{As Fig.~\ref{ansatz3 phantom}, but showing the functions 
$X(a)$, $F(a)$, and $F(X)$. 
}
  \label{ansatz3 2 phantom}
\end{figure}

We explore three different cases, corresponding to diverse physical 
situations.  The example with $x=3$ describes well a thawing equation 
of state, evolving away from cosmological constant behavior, but would 
be an unstable k-essence model.  For $x=-3$, the dynamics is freezing, 
approaching a cosmological constant, and the corresponding k-essence 
model has positive sound speed squared for all redshifts.  The 
model with $x=-6$ and $w_0=-1.1$ gives a toy phantom model which 
approaches a cosmological constant, and lies in the stable region 
only at recent redshifts.  These cases are illustrated in 
Figs.~\ref{ansatz3}-\ref{ansatz3 2 phantom}.

\subsection{Generalized Chaplygin Gas} 
\label{subsec:4}

Dark energy that behaves like matter ($w = 0$) at early times and like 
a cosmological constant ($w = -1$) at late times can be described by an 
equation of state
\beq
\label{}
1 + w(a) = \left[1 - \frac{w_0}{1+w_0}a^{3 (n+1)}\right]^{-1},
\eeq 
where we take $w_0 \in (-1,0)$ and $n > 0$.  
A model that produces exactly this equation of state is the generalized 
Chaplygin gas (GCG, \cite{BenBerSen02}), which can also be described by 
a perfect fluid/gas with equation of state $p = -A/\rho^{n}$, 
where $A$ is a constant. The original model, the Chaplygin gas (\cite{Chap1904,KamMosPas01}), corresponds to the case $n = 1$. The GCG has been extensively 
studied in hope of providing a unified description of dark matter and dark 
energy, but in such an approach it has problematic issues with structure 
formation \cite{Sandetal04,BeanDore03,AmWagFin05}.  However, it is still interesting to 
study as a dark energy model.  
This equation of state can be reproduced by a 
k-essence model (see, e.g., \cite{BenBerSen02}) and, of the dynamical 
equations of state we discuss in this paper, this is the only one where an 
{\it explicit\/} expression can be found for the k-essence Lagrangian 
density $F(X)$. The Lagrangian for $n = 1$ can be
linked to the tachyon in string theory and to the dynamics of branes
(as referred to in \S\ref{sec:motiv}). 

\begin{figure}
  \begin{center}{
  \includegraphics*[width=8.5cm]{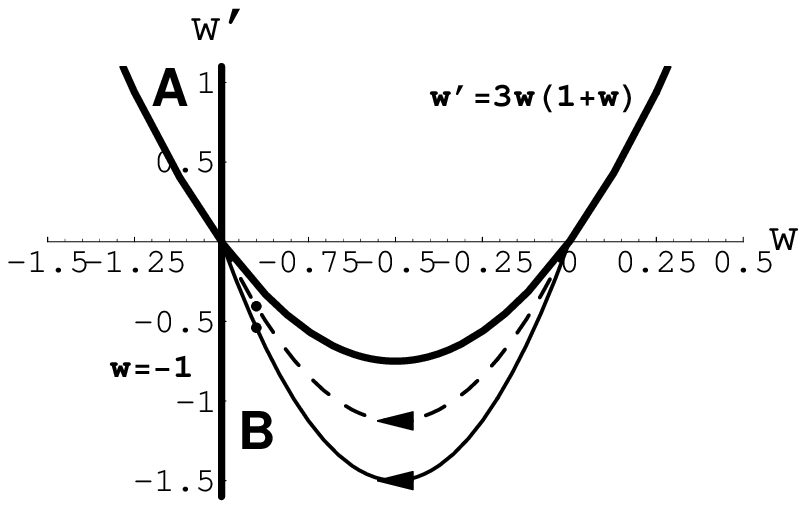}
  \includegraphics*[width=8.5cm]{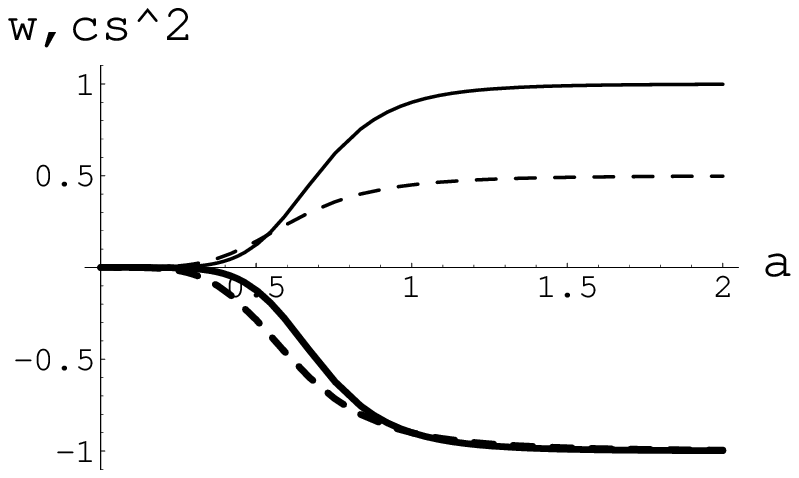}
  }
  \end{center}
  \caption{Ansatz D (the generalized Chaplygin gas) with $w_0 = -0.9$ 
and $n = 1$ (solid) or $n = 0.5$ (dashed). The first panel shows the 
trajectories in the $w$-$w'$ plane.  
The direction of increasing scale factor is indicated by arrows and 
dots mark the values of $w$ and $w'$ today.  Since the equations 
of state lie completely in region B, they can be obtained from purely 
kinetic k-essence 
Lagrangians (see Fig.~\ref{ansatz4 2}). The second panel shows the 
equation of state $w$ (thick lines) and the adiabatic sound speed squared $c_s^2$ (thin lines) as a function of scale factor $a$.}
  \label{ansatz4}
\end{figure}

From the expression for $w(a)$ one finds that its dynamical trajectory 
is a parabola given by $w' = 3 (n + 1) w (1 + w)$, with $w(a)$ evolving 
from $0$ to $-1$ (also see the mocker model in \cite{Linder06}).  
For $n>0$ this equation of state 
always lies completely in the allowed region B. It is therefore possible 
to reproduce it with a stable kinetic k-essence solution. Note that in 
the limit $n \to 0$ we approach the boundary of the allowed regions, the 
constant pressure line.  (For $n>1$ there will be epochs where the 
trajectory crosses the null line given by Eq.~(\ref{null}); see 
\S\ref{sec:stability}.)  We consider two examples ($n = 0.5, 1$ with 
$w_0=-0.9$) in Fig.~\ref{ansatz4}.

\begin{figure}
  \begin{center}{
  \includegraphics*[width=8.5cm]{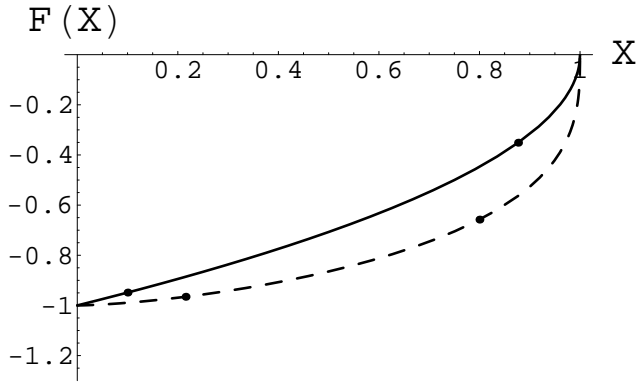}
  }
  \end{center}
  \caption{As Fig.~\ref{ansatz4}, but plotting the corresponding k-essence 
Lagrangian density $F(X)$ (with $F$ in units of $A^{1/(n+1)}$).  
$X$ starts at one and then moves to zero as the scale factor increases. 
Markers indicate the points where $z = 1$ (right) and $z = 0$ (left).}
  \label{ansatz4 2}
\end{figure}

To construct $F(X)$, we first solve equation (\ref{diff X}) to find
\begin{equation}
\label{}
X(a) = \left(1 - \frac{w_0}{1+w_0}a^{3(n+1)}\right)^{-\frac{2n}{n+1}},
\end{equation}
where we have chosen the normalization such that $X$ goes from $1$ at $a = 0$ to $0$ at $a = \infty$. The expression can easily be inverted to give $a(X)$. Subsequently, Eq.~(\ref{F using X}) leads to 
\beq
\label{F GCG}
F(X) = - A^{\frac{1}{n+1}} (1 - X^{\frac{n+1}{2n}})^{\frac{n}{n+1}}
\eeq
(cf. \cite{BenBerSen02}), where $A$ is the constant appearing in the GCG equation $p = -A/\rho^n$. We plot two examples in Fig.~\ref{ansatz4 2}.

\section{Conclusions}
\label{sec:conclusions}

Kinetic k-essence is in some sense an equally probable solution to 
the dark energy conundrum as quintessence, trading a single potential 
function $V(\phi)$ for a single kinetic function $F(X)$.  Similarly, 
one can find equivalent motivations for it from quantum field and 
extra dimension theories. 

We have established stability regions for such models within the 
equation of state phase space, based on the necessary condition of 
a non-imaginary sound speed.  Conversely, we find closed form solutions, 
given some equation of state $w(a)$, for a dynamically corresponding 
kinetic k-essence Lagrangian $F(X)$.  
Eqs.~(\ref{fconditions1}-\ref{fconditions3}) and (\ref{fxxp}-\ref{fxxm}) 
constitute a simple prescription -- ``by eye'' -- for characterizing 
kinetic k-essence theories. 

Using these results, we investigated the limits to dynamical degeneracy 
between kinetic k-essence and quintessence.  
Analyzing four types of equations of state representing diverse dynamics, 
we found the limits in redshift defining the degeneracy region.  
For several equation of state models, both $w>-1$ and phantom, this 
implies that one could rule 
out all kinetic k-essence models with a sufficient redshift range of 
measurements.  On the other hand, an equation of state similar in form 
to the generalized Chaplygin gas could be equally described by k-essence 
for all redshifts (as was already known).  This clear definition of degeneracy 
regions offers increased hope that with future observational data on 
the dark energy dynamical and microphysical effects we can discern 
which approach describes the new physics behind our accelerating universe.

\section*{Acknowledgments} 

This work has been supported in part by the Director, Office of Science,
Department of Energy under grant DE-AC02-05CH11231. 

\bibliography{k_essence}

\section*{Appendix}

As an example, consider the simple case of a $3$-brane embedded in a $4+1$ dimensional space-time with a fixed Minkowski metric $\eta_{i j}$, $i, j = 0,\dots 4$. The brane can be parametrized by four world-sheet coordinates $x^{\mu}$ ($\mu = 0,\dots 3$) so that the location of the brane in space-time (or target space) is given by $X^i = X^i(x^{\mu})$ ($i=0,\dots 4$). In general, the Nambu-Goto action of a $p$-brane is given by the volume of its world-sheet
\beq
\label{}
S_{\rm NG} = \int d^4x \sqrt{- {\rm det}(g_{\mu \nu})},
\eeq
where $G_{i j}$ is the target space metric and $g_{\mu \nu} = \pa_{\mu} X^i \pa_{\nu} X^j G_{i j}$ is the induced metric on the brane. In our example, $G_{i j} = \eta_{i j}$.

In the static gauge, the world-sheet coordinates are chosen to coincide with the first four target space coordinates: $x^{\mu} = X^{\mu}$ for $\mu = 0,1,2,3$. We now call the fifth target space coordinate $X^4 = \phi$ to emphasize that it is a scalar from the world-sheet point of view. In this parametrization/gauge, $g_{\mu \nu} = \eta_{\mu \nu} - \pa_{\mu} \phi \pa_{\nu} \phi$. If we now assume that $\phi$ only depends on the time coordinate $t = x^0$, we get the Lagrangian
\beq
\label{}
S = \int d^4 x \sqrt{1 - 2X}.
\eeq
This is equivalent to the Chaplygin gas Lagrangian (see also \S\ref{subsec:4}) if there we assume a fixed Minkowski background. We can reproduce Lagrangians of the more general form $\mathcal{L} = - V(\phi) \sqrt{1 - 2 X}$, with a variety of $3+1$ dimensional backgrounds, by having different $4+1$ dimensional backgrounds $G_{i j}$. The ``potential'' $V(\phi)$ then corresponds to a warp factor in the metric.

We wish to stress that this discussion is merely meant to show that there are physical ways to get k-essence Lagrangians. We do not claim that the above constitutes a realistic cosmological model.

\end{document}